\documentclass[AMS,Times1COL]{USG} 

\usepackage{algorithm}
\usepackage{amsmath}
\usepackage{cleveref}
\usepackage{graphicx}
\usepackage{multicol}
\usepackage{xcolor}
\newcommand{\edit}[1]{\textcolor{black}{#1}}

\articletype{Original Article}%

\received{Date Month Year}
\revised{Date Month Year}
\accepted{Date Month Year}
\journal{Journal}
\volume{00}
\copyyear{2026}
\startpage{1}

\begin{document}

\title{kobra: a new Vlasov code intended for plasma-wall modeling}

\author[1]{Sebastian Konewko}

\author[1,2]{Nathan Maestracci}

\author[1]{Sven Van Loo}

\authormark{Konewko \textsc{et al.}}
\titlemark{kobra: a new Vlasov code for plasma wall modeling}

\address[1]{\orgdiv{Department of Applied Physics}, \orgname{Ghent University}, \orgaddress{\state{Ghent}, \country{Belgium}}}

\address[2]{\orgdiv{Insitut Jean Lamour}, \orgname{Universit\'e de Lorraine}, \orgaddress{\state{Nancy}, \country{France}}}

\corres{Sebastian Konewko, Department of Applied Physics, Ghent University, Ghent 9000 ,Belgium, \email{sebastian.konewko@ugent.be}}


\abstract[Abstract]{In a fusion device plasma-wall interactions \edit{on the sheath scale} can be modeled as a collisionless problem. When modeling these regions  particle-in-cell codes suffer from statistical error originating from undersampling the velocity space. On the other hand, Vlasov codes do not have this issue as they evolve the full distribution function. Here, we present a new finite-volume Vlasov code, \textit{kobra}, equipped with adaptive-mesh refinement to reduce computational effort. Currently, the code solves the Vlasov-Poisson equations. We validate our code \edit{in 1d1v and 1d2v} using established benchmarks, i.e. the two-stream instability, Landau damping, \edit{the Dory-Guest-Harris instability}, and also a classical electrostatic plasma sheath. We find that the code reproduces the theoretical properties of these problems well. \edit{More importantly, the adaptive grid provides a computational gain that is likely to scale to higher dimensional, plasma-wall simulations.}
}

\keywords{Vlasov-Poisson, Kinetic modeling, plasma-wall interaction, finite volume, AMR}
\articledoi{}

\maketitle



\section{Introduction}
\label{sec:Introduction}

For the operation of a fusion tokamak, it is fundamental to understand the interaction of the plasma with the reactor wall and especially the effect on the plasma-facing 
components of the device. Fluid codes are used successfully to model the scrape-off layer (SOL) with realistic magnetic equilibria and wall geometry and with relatively 
modest computing requirements. However, they cannot self-consistently treat the kinetic effects important in the SOL \edit{and the sheath-scale physics which set the 
boundary conditions for the simulations}. Therefore, a kinetic approach is needed to fully capture the plasma's behavior \cite{TSKHAKAYA2021100893}.

Typically, a particle-in-cell (PIC) code is the most straightforward approach as it follow the trajectories of individual particles while calculating the 
electromagnetic fields on a grid level. However, this approach suffers from statistical noise when undersampling particles in \edit{regions with large density 
gradients, such as the transition from the pedestal to the SOL.} In contrast, solving the Vlasov equation allows one to capture the full velocity distribution, 
eliminating the statistical errors. However, this gain comes at a computational cost \edit{as discretizing the six dimensional phase space $(\textbf{x}, 
\textbf{v})$ generates} an extraordinary need for memory and cpu time. Also, finite-volume codes suffer from numerical diffusion. \edit{Here, we introduce 
\textit{kobra}, a new finite-volume Vlasov-Poisson code with adaptive-mesh refinement (AMR). Using AMR techniques reduces the number of computational grid 
cells and allows to address hitherto impossible computational models.}

\edit{While \textit{kobra} is being developed for near-wall plasma studies, we first start with a general purpose code for collisionless, electrostatic problems. 
This simplifies the development as we can focus on the numerical algorithms solving the Vlasov and Poisson equations while simultaneously incorporating AMR (\Cref{sec:kobra}). 
The complexities involved in near-wall plasma modeling such as, e.g. particle collisions, wall sputtering and erosion, recombination of ions and electrons, and secondary emission 
of electrons from the wall, do not affect the underlying algorithms and will be included in the future. Also, straightforward comparison can be made between the numerical 
and theoretical results of \edit{1d1v and 1d2v} benchmark problems necessary for code validation (\Cref{sec:Benchmarks}).} Then, we discuss the setup and results of simulations of a plasma sheath in \Cref{sec:PlasmaSheath}. Finally, concluding remarks are made in \Cref{sec:Conclusion}.

\section{kobra}
\label{sec:kobra}

To model a collisionless plasma kinetically, we need to solve the Vlasov equation:
\begin{equation}
    \frac{\partial f_s}{\partial t} + \textbf{v} \frac{\partial f_s}{\partial \textbf{x}} + \frac{\textbf{F}_s}{m_s} \frac{\partial f_s}{\partial \textbf{v}} = 0,
    \label{eq:Vlasov}
\end{equation}
given the Lorenz force $\textbf{F}_s = q_s( \textbf{v}\times\textbf{B} + \textbf{E})$, where $f_s(\textbf{x},\textbf{v},t)$ is the distribution function in phase space, and the subscript $s$ denotes the species. Assuming electrostatic plasmas, we ignore the self-generated magnetic field and interpret $\textbf{B}$ as a constant magnetic field externally imposed upon the plasma. Therefore, we must only solve for the self-generated electric field $\textbf{E}$ which can be written in terms of a potential scalar field $\textbf{E} = - \nabla \phi$. We solve the Poisson equation:
\begin{equation}
   \nabla^2 \phi = -\sum_s q_s \ n_s(\textbf{x}).
   \label{eq:Poisson}
\end{equation}
where $n_s(\textbf{x}) = \int f_s(\textbf{x,v})d\textbf{v}$ are the particle number densities. Equations \ref{eq:Vlasov} and \ref{eq:Poisson} form the Vlasov-Poisson system. 
\edit{For simplicity we non-dimensionlize this system so that a unity of length is scaled with the Debye length ($\lambda_D$), velocity with the thermal electron velocity ($v_{Te}$) and time inversely with the plasma frequency ($\omega^{-1}_p$).}

We have two types of equations to solve: the hyperbolic Vlasov equation, and the elliptic Poisson equation. Using a finite-volume method this requires two different approaches. \edit{For the Vlasov equation which is an advection equation we use a second-order Runge-Kutta scheme and a second-order upwind scheme with a slope limiter to avoid spurious oscillations in regions with large gradients (see e.g.  \cite{Olivieri}). As this only requires to calculate the fluxes at cell interfaces, it can easily be extended to AMR grids. Here we adopt the hierarchical adaptive gridding strategy described in \cite{FalleGiddings} by using a sequence of grids $G_0 \ldots G_N$, such that, if the mesh spacing on $G_0$ is $\Delta x$, then it is $\Delta x/2^N$ on $G_N$. Grids $G_0$ and $G_1$ are then used to cover the whole domain of interest, but $G_2  \dots G_N$ only exist at those locations in the flow where the nature of the solution demands it. Solutions of the Vlasov equation are also calculated on each grid taking into account flux correction across levels. The refinement process is controlled by using these solutions to obtain a cell-by-cell estimate of the truncation error at the end of each time step. While AMR techniques typically use a different time step for each grid level, we adopt a synchronous time step on each level for simplicity.}

\edit{Determining the electrostatic field requires the solution of a sparse linear algebraic system. A robust way of accelerating the usual Jacobi or Gauss-Seidel iterative methods involves multigrid methods which correct the solution by solving a coarser problem. The choice for a multigrid solver is motivated by its natural extension to AMR grids through the Full Approximation Scheme (FAS) and is easily parallelised \cite{Brandt}. As the smoothing operator in the FAS we use lexicographic Gauss-Seidel. Note that the Poisson equation only needs to be solved in coordinate space with velocity-integrated number densities. Therefore, it is more convenient to use an auxiliary grid in coordinate space only which reflects the AMR structure of the full phase-space. We also implement velocity-reduction algorithms that maintain spatial fidelity and accuracy while integrating across the velocity dimensions to determine the particle number densities.}

\edit{When modeling multi-species plasmas the simulation needs to track the phase-space distribution function for each species. As there is a large range of particle masses and thermal velocities, a common practice is to choose a different computational domain for the velocity space of each species, and taking larger grids for the low-mass species compared to the high-mass ones (e.g. electrons versus protons). Otherwise, if a single grid is used, the high resolution needed for the massive particles prohibitively reduces the computational time step proportional to the electron dynamical time. However, these multiple grids require an increased memory requirement. Alternatively, it is possible to apply an independent velocity-scaling to each species (relative to the global velocity scaling applied to non-dimensionalize the Vlasov equation) while keeping a single phase-space grid. This combines the advantages of both a single phase-space grid with individual velocity domains for each species and is used here.}


\section{Benchmarks}
\label{sec:Benchmarks}
To test and benchmark \textit{kobra} we \edit{first apply it to the 1d1v Landau damping and two-stream instability problems \cite{VogThesis,CHENG1976330,POHN200581}.} In the following simulations, we use periodic boundary conditions for $x$ and free-flow boundary conditions in $v$. Furthermore, we assume static ions with $n_i(x)= 1$ and set ${\bf B} = 0$. The initial electron distribution is given by
\begin{equation}
    f_0 (x,v) = \frac{1}{\sqrt{2 \pi}} \left[ 1 - \alpha \cos \left(\frac{x}{2}\right)\right] \exp \left[ \frac{-v^2}{2} \right], \quad {\rm and} \quad 
     f_0 (x,v) = \frac{v^2}{\sqrt{2 \pi}} \left[ 1+ \alpha \cos \left(\frac{x}{2}\right)\right] \exp \left[ \frac{-v^2}{2} \right].
    \label{eq:ldamping}
\end{equation}
for Landau damping and  the two-stream instability, respectively. 
For both cases, the amplitude of the perturbation $\alpha$ was chosen to match work done by \cite{VogThesis}, i.e. $\alpha = 0.5$ for strong Landau damping and $\alpha = 0.01$ for the two-stream instability. The numerical domain for both models is given by $(x,v) \in [0,4 \pi] \times [-10,10]$ with a  default resolution of 256$\times$256.

\begin{figure}[h!]
    \centering
     \includegraphics[scale=0.58]{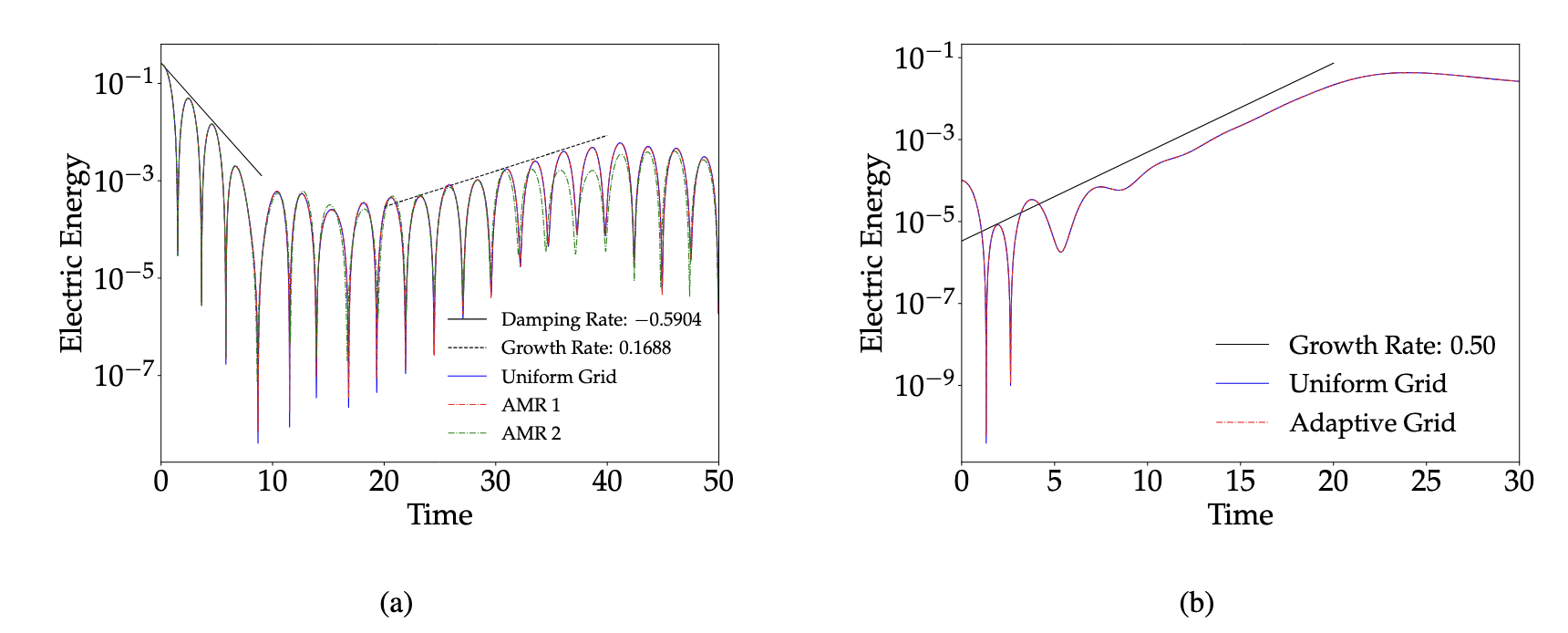}
        \label{fig:Ldamping}
        \label{fig:Twostream}

    \caption{\edit{(a) Temporal evolution of the electric energy for strong Landau damping for a uniform (blue solid) and AMR with refinement tolerance of $10^{-3}$ (red dashed; AMR1) and $5\times 10^{-3}$ (green dashed; AMR2) grid. The black lines show the decay (solid) and growth rates (dashed) from \cite{VogThesis}. (b) Same as (a) but for the two-stream instability. The solid black line shows the linear theory growth rate.}}
    \label{fig:Fig1}
\end{figure}

\edit{\Cref{fig:Fig1} shows the evolution of the electric field energy for both problems with uniform grids. In \Cref{fig:Fig1}a we see, for the strong Landau damping, the initial damping of the electric field ($0 < t < 15$) as energy is transferred from the plasma wave to the electrons. Then, for $15 <  t < 40$, there is an inversion as now energy is transferred back to the plasma wave which becomes unstable. Finally, the electric energy saturates due to particle trapping near $t=40$.  Both the damping and growth rate of the electric energy are the same as in \cite{VogThesis}. For the two-stream instability, counterstreaming electron beams excite a plasma wave which then grows exponentially until it saturates around $t = 20$ as can be seen in \Cref{fig:Fig1}b. The exponential growth rate also follows the predicted theoretical one. }

\edit{Using AMR the dynamical evolution remains unchanged (see \Cref{fig:Fig1}), while reducing the computational cost as only regions with large truncation errors are refined. For refinement above a truncation error of $10^{-3}$, the AMR simulations are 1.7 and 2.3 (Landau damping and two-stream instability, resp.) times faster and use about 1.75 times less memory than the uniform runs.} Note that the computational gain depends on the AMR parameters, i.e. the number of grid levels and refinement tolerance. \edit{For a higher tolerance 
of the truncation error, less grid cells are refined. Hence, the simulation runs faster, e.g., for a refinement tolerance of  $5\times10^{-3}$ the Landau damping model is 2.8 times faster than the uniform model. However, this also degrades the accuracy, as can be seen in \Cref{fig:Fig1}a at $t\approx 35-40$. Numerical diffusion prevents the total energy to be conserved to machine precision and, as this is proportional to the grid spacing, coarser grid cells produce a larger error. Therefore, it is imperative to balance the computational gain with numerical accuracy. We did not investigate the optimal settings for the AMR as they are likely problem dependent. Note that, for the uniform runs, the total energy also is not conserved. However, the relative error remains below $10^{-3}$ even after $t=100$ and, thus, remains acceptable as they produce the correct dynamical behavior.}

\begin{figure}
    \centering
     \includegraphics[scale=0.58]{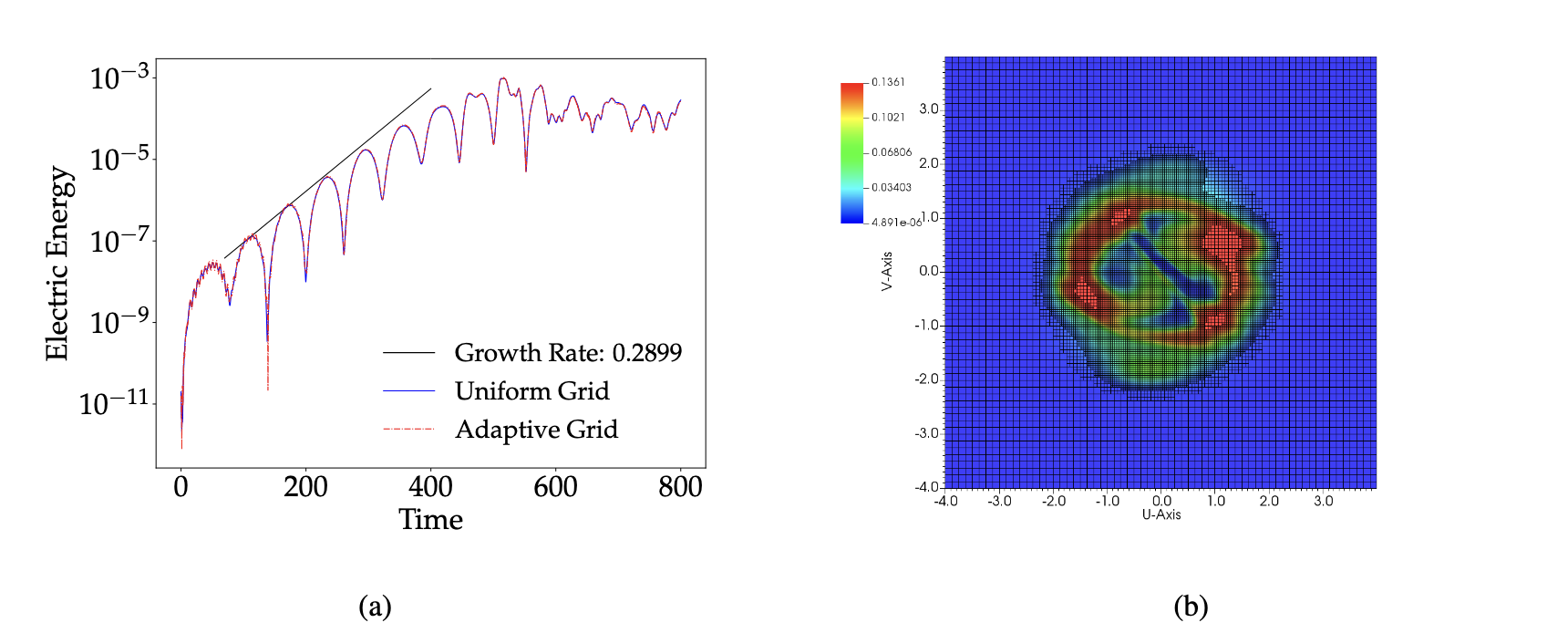}
        \caption{\edit{(a) The evolution of the electric energy for the DGH instability. (b) The electron distribution at $x = 0$ in velocity space when the instability has saturated ($t = 500$) with AMR grid overlaid.}} 
    \label{fig:DGHdouble}
\end{figure}

\edit{Our final test problem is the Dory-Harris-Guest (DGH) instability which a 1d2v cyclotron harmonic instability. It allows the benchmarking of magnetized plasmas in higher, but limited, dimensionality. The magnetic field is in the $\hat{z}$ direction and normalized by the ratio of the cyclotron and plasma frequencies of the electrons, i.e. $\Omega_c  = \omega_c/\omega_p$. Following \cite{VOGMAN2014101}, the initial condition is given by:
\begin{equation}
f_0(x,v_x,v_y) = \frac{j}{2\pi j!} \left(\frac{v_x^2 + v_y^2}{\alpha_\perp^2} \right)^j \exp\left(\frac{-j}{2}(v_x^2 + v_y^2) \right) \left(1 + \epsilon \sin \left( 4\theta - \frac{\tilde{k} \ \Omega_c}{\sqrt{2}}x \right) \right)
    \label{eq:DGH}
\end{equation}
where $\theta = \arctan \left(v_y/v_x\right)$. The numerical domain is $(x,\textbf{v})\in [0,L=2\pi/\tilde{k}] \times [-4,4]^2$ with a resolution of $(128,256^2)$. We set $j=6$, $\tilde{k} = 4.65$ and $\omega_p/\omega_c = 20$, and perturb our distribution with $\epsilon = 10^{-3}$. For these parameters the DGH equilibrium distribution both oscillates and becomes unstable with a linear growth phase between $138 < t < 510$ after which it saturates. The theoretical growth rate is 0.2899 with an oscillation frequency of 1.0361. Indeed, \Cref{fig:DGHdouble}a shows this behavior with a numerical growth rate of 0.2635 and oscillation frequency of 1.030 with again no difference between the uniform and AMR simulations. However, the memory requirements for the AMR simulations are significantly reduced from 11.4Gb to 2.7Gb (a factor of 4.2) while a speed-up of a factor 3.8 (9.5hrs compared to 37hrs) is achieved. \Cref{fig:DGHdouble}b shows the grid levels on top of the electron distribution at $x=0$ in velocity space when the instability is near saturation (at $t=500$) to illustrate that only a small part of the numerical domain is covered by the finest grid.}

\section{Plasma Sheath}
\label{sec:PlasmaSheath}
As we seek to model the plasma-wall interaction in fusion devices, we look at the seminal paper by \cite{Chodura} studying the plasma incident on a wall with an external oblique magnetic field acting upon the plasma. While this requires a 1d3v model, we first start with the electrostatic plasma sheath in 1d1v and, thus, neglect the magnetic field. \edit{We note that this plasma sheath setup is a specific case of the sheaths discussed in \cite{10.1063/1.859279} and therefore the steady state can be solved analytically.}  

The distribution functions of the ions and electrons streaming in (i.e. positive velocities) from a charge-neutral plasma side at $x = 0$ are set as fixed boundary conditions. These are given by 
\begin{equation}
    f_{e}(0,v>0) = \frac{n_{e0}}{\sqrt{2\pi}} \exp \left( -\frac{v^2}{2} \right),
    \quad {\rm and} \quad 
    f_{i}(0,v>0) = n_{i0} \  \delta(v-\textbf{v}_{0}),
\end{equation}
where for the ions where we assume 'cold' ions with a single inflow velocity $v_0$. This inflow velocity needs to satisfy the Bohm limit $\textbf{v}_{0} \geq \sqrt{1/m_i}$ (with $m_i = 1836$ in normalized units) for a plasma sheath to form. Here we use $v_0 = 0.2$. For the negative velocities, the boundary condition at $x=0$ is free flow, i.e. $\partial f_{i,e}/\partial x = 0$. Furthermore, $n_{i0}=1$ and $n_{e0} = 1.115$. These values are different to ensure that, at equilibrium, the plasma is charge neutral near $x=0$ and thus also $E_x(x=0) = 0$. 
At the other side of the numerical domain ($x=10$), we assume an absorbing wall, i.e. free-flow for the positive velocities and a zero-inflow condition for the negative velocities. Furthermore, we allow the wall to be charged and, thus, use a floating boundary potential, which is essentially a Neumann boundary condition,  when solving for the Poisson equation. \edit{The potential at the plasma side also uses a Neumann boundary condition. After every time step the potential is recalibrated so that, at the plasma side, $\phi = 0$. This approach allows the plasma to connect freely with the quasi-neutral core and is more self consistent than imposing the Dirichlet condition $\phi = 0$ directly.} Furthermore the numerical domain is initially empty and is given by $(x,v)\in[0, 10] \ \text{x} \ [-7.5,7.5]$ with an effective resolution of $(128,384)$. \edit{We scale the ions such that $-0.75 \leq v \leq 0.75$ while maintaining the same resolution as the electrons}.

\begin{figure}[h!]
    \centering
     \includegraphics[scale=0.58]{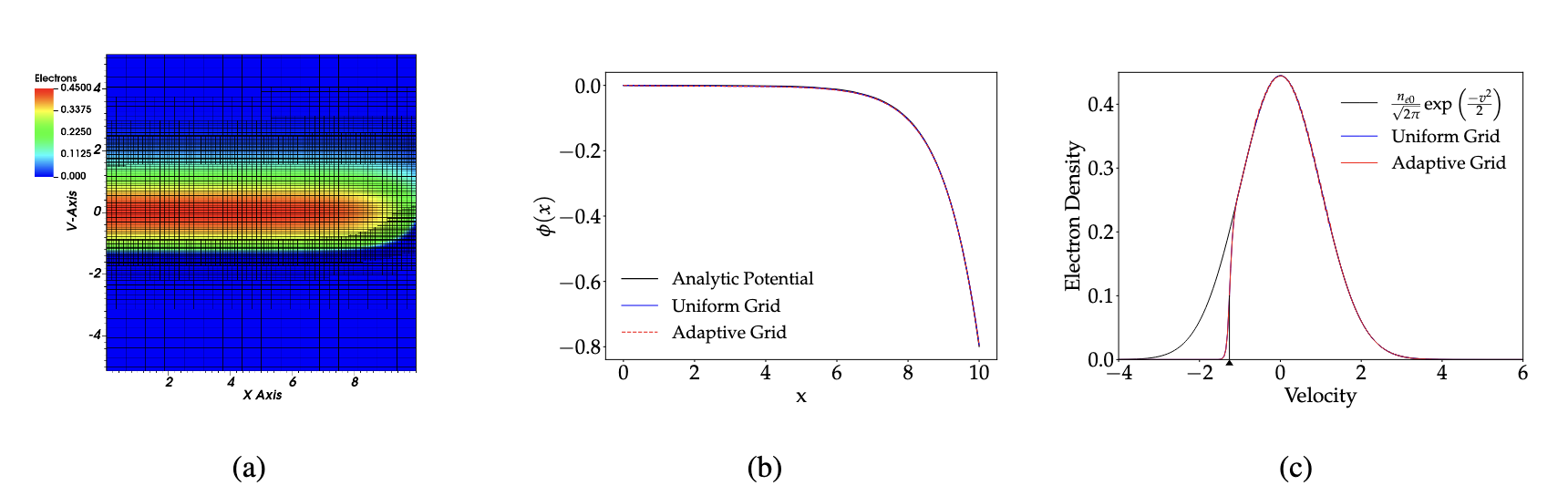}
     \caption{\edit{(a) Steady-state electron distribution in phase-space with AMR grid overlaid. (b) Converged potential distribution of the uniform and AMR grids compared to the analytic solution. (c) Steady-state electron distribution function in velocity space at $x=0$ for the uniform and AMR grids. The black line represents a full Maxwellian distribution at the bulk plasma boundary with the vertical line indicating the truncation velocity at $v=-1.26$.}}
    \label{fig:Fig3}
\end{figure}



\edit{As plasma starts to flow in from the plasma side, electrons reach the wall before the ions.}
\edit{Eventually, the ions reach the wall as well and a steady-state is attained with a positively-charged layer (i.e. the plasma sheath) in front of the negatively-charged wall. The electron distribution of the plasma sheath is shown in \Cref{fig:Fig3}a.}  \edit{For this steady-state sheath, a number of analytic properties can be compared with the numerical ones. For example, the wall potential is dependent only on $v_0$ as 
$\exp(\phi_{\rm wall})/\left[1+ {\rm erf}\left(\sqrt{-\phi_{\rm wall}}\right)\right] = \sqrt{\frac{\pi}{2}}v_0$. Hence, $\phi_{\rm wall} = 0.739$ for our parameters. Then, from
$\phi_{\rm wall}$ the spatial distribution $\phi(x)$ can be derived. \Cref{fig:Fig3}b shows this theoretical $\phi(x)$ compared to the numerical obtained one. We see that the distributions match almost identically. Herefrom, the electron distribution can also be reconstructed at every position $x$. In effect the distribution is a truncated Maxwellian where the cut-off velocity is given by $v_{\rm cutoff}(x) = - \sqrt{2 (\phi(x) - \phi_{\rm wall})}$. This is because the potential decelerates electrons streaming in from the plasma-side and only electrons with enough kinetic energy (or $v > -v_{\rm cutoff}(x))$ reach the wall, while the others are reflected and form the negative velocity part of the distribution. \Cref{fig:Fig3}c shows the electron distribution at $x=0$ and the simulations reproduce the truncation accurately although there is some smoothing because of the discretization.} 

\edit{While the uniform and AMR simulations provide similar results, the computational requirements needed for the AMR are, as in the benchmark problems, much lower than for the uniform grid. The latter run took around 3 hours with 50 Mb, while the former finished in 22 minutes (about 8 times faster) and used half of the memory. \Cref{fig:Fig3}a shows that, in order to adequately model the plasma sheath, it is necessary to capture the truncation velocity in the electron distribution (the finest grid in the negative velocities) and the ion delta function distribution located at $v=0.2$. Given the ion velocity scale of $10v_{T,e}$ this is shown in the figure near $v \approx 2$ in the positive velocities of the phase-space. The rest of the distribution can be well described by a lower resolution (see \Cref{fig:Fig3}c). }



\section{Discussion and Conclusions}
\label{sec:Conclusion}
\edit{In this paper we have presented the newly-developed finite-volume Vlasov-Poisson code \textit{kobra}. While \textit{kobra} is eventually intended for plasma-wall modeling, 
we defer the implementation of the near-wall plasma physics and focus on the basic algorithms governing the Vlasov-Poisson system. As finite-volume Vlasov codes are typically
constrained by the computational scalability to the full 3D-3V phase space, it is necessary that techniques are used to reduce the memory requirements. Here we implement AMR which
only uses high-resolution in regions with large truncation errors and is compatible with advection and multigrid methods (for the Vlasov and Poisson equation, respectively). Applying 
\textit{kobra} to benchmark problems in 1d1v and 1d2v (magnetized) plasmas, we show that the implemented algorithms perform well and reproduce the expected theoretical behavior both 
for uniform and AMR grids. Furthermore, AMR significantly reduces the memory requirements of the models and produces an associated computational speed-up. There is 
some indication that, for higher dimensions, a larger reduction is achieved as more of the numerical domain can be covered by coarser grids. While this is promising, it still needs to 
be confirmed by higher-dimensional models. Also, specific values for the improvement are dependent on the AMR parameters used. For example, higher values for the refinement tolerance 
reduce the needed memory. However, note that these runs are also less accurate as finite-volume codes suffer from numerical diffusion leading to errors in energy conservation. Thus, 
the use of AMR requires a careful balance between computational gain and accuracy.}

\edit{As we eventually look to model tokamak edge physics, we already apply \textit{kobra} to a collisionless electrostatic sheath problem modeling both ions and electrons. Again the 
analytic properties of the sheath are well reproduced by both uniform and AMR models. Here the gain due to AMR is again apparent and even larger than for the benchmark problems. 
High resolution is only required in a very localised region of velocity space, i.e. at the electron truncation velocity and around the ion delta distribution. This suggest that,
for a 1d3v model, the computational gain can be even larger. This will be examined in a subsequent paper modeling the magnetised plasma sheath of 
\cite{Chodura}.}

\edit{While these early results for \textit{kobra} are encouraging, it is necessary to further reduce in computational costs. This can be achieved by implementing 
higher-order schemes and anisotropic AMR. The former provides a higher accuracy for lower resolution, while the latter only refines the grid in specific directions to capture rapid,
directional changes in physical solutions. Simultaneously we  will include additional physics, such as collisional operators to model the particle interactions and move from collisionless sheath physics to plasma wall interactions}.

\bmsection*{Acknowledgments}
SK and SVL gratefully thank the BOF grant funding of Ghent University (BOF/STA/202202/015). NM thanks the Erasmus$+$ funding programme to support an internship research position at the Infusion group of Ghent University. VisIt, a parallel data visualization tool developed at the Lawrence Livermore National Laboratory (LLNL), was used to generate plots \cite{HPV:VisIt}.

\bmsection*{Conflict of interest}
The authors have stated explicitly that there are no conflicts of interest in connection with this article

\bmsection*{DATA AVAILABILITY STATEMENT}
The data that support the findings of this study are available from the corresponding author upon reasonable request.

\bibliography{PETSeb}

\end{document}